\documentclass{elsart}

\usepackage[dvips]{graphicx}
\usepackage{array}

\usepackage{amssymb}

\begin{document}

\begin{frontmatter}



\title{Spatial Characteristics of Joint Application Networks in Japanese Patents\thanksref{1}}


\author[a1]{Hiroyasu Inoue\corauthref{cor1}},
\ead{hirinoue@doshisha-u.jp}
\author[a2]{Wataru Souma},
\author[a3]{Schumpeter Tamada}

\address[a1]{ITEC, Doshisha University, Kyoto 602-8580, Japan}
\address[a2]{NiCT/ATR CIS Applied Network Science Laboratory, Kyoto 619-0288, Japan}
\address[a3]{Institute of Business and Accounting, Kwansei Gakuin University, Hyogo 662-8501, Japan}

\corauth[cor1]{Corresponding author. Tel: +81-75-251-3779. Fax: +81-75-251-3139.}

\thanks[1]{This paper is an output of the research project,
Globalization and Vertical Specialization:
A four country comparison, supported by Doshisha University's
ITEC 21st Century COE (Centre of Excellence) Program
(Synthetic Studies on Technology, Enterprise and Competitiveness Project).}

\begin{abstract}
Technological innovation has extensively been studied
to make firms sustainable and more competitive.
Within this context,
the most important recent issue has been
the dynamics of collaborative innovation among firms.
We therefore investigated a patent network,
especially focusing on its spatial characteristics.
The results can be summarized as follows.
(1) The degree distribution in a patent network
follows a power law.
A firm can then be connected to many firms
via hubs connected to the firm.
(2) The neighbors' average degree has a null correlation,
but the clustering coefficient has a negative correlation.
The latter means that there is a hierarchical structure
and bridging different modules may shorten the paths between the nodes in them.
(3) The distance of links
not only indicates the regional accumulations of firms,
but the importance of time it takes to travel,
which plays a key role in creating links.
(4) The ratio of internal links in cities
indicates that 
we have to
consider the existing links firms have
to facilitate the creation of new links.
\end{abstract}

\end{frontmatter}

\section{Introduction}
\label{sec:int}

Technological innovation has extensively been studied
to make firms sustainable and more competitive.
The most important recent issue has been
the dynamics of collaborative innovation among firms \cite{Porter98}.
Moreover,
a lot of countries are promoting industrial cluster policies
that facilitate collaborative innovation among firms
in specific regions,
and emphasizing that the key is creating networks among firms.
However, 
studying industrial clusters based on networks of firms
is not sufficient
because of the difficulty of obtaining comprehensive linkage data.

There have been numerous extensive studies on innovation
in the social science
based on networks \cite{Powell06}.
However,
these studies have focused on the details of specific collaborative innovations,
and have only treated one thousand firms at most.
All regional firms and their networks should be studied
to enable industrial clusters to be discussed,
and the firms should number more than ten thousand.

This paper focuses on networks generated by
joint applications of patents in Japan.
These networks can cover all Japanese firms,
and these enable us to study industrial clusters.
Joint applications are common in Japan,
even though they are not popular in Europe or the United States.
This is why there have been no similar studies in those areas.

The entire dynamics of collaborative innovation
cannot be observed by focusing on the joint applications of patents.
This is because all innovation is not revealed in patents,
and all patents cannot lead to innovation.
However, this problem can be ignored.
Since exact distinctions, whether various patents
have contributed to innovation do not matter,
we pay attention to the structure
of innovation network among firms using the patent network.

This paper is organized as follows.
Section 2 explains the patent data discussed in this paper
and joint application networks derived from them.
In Section 3, we discuss spatial characteristics
that are important for industrial clusters, and conclude this paper.

\section{Japanese patent data and joint application networks}
\label{sec:data}

The Japanese Patent Office publishes patent gazettes,
which are called
{\it Kokai Tokkyo Koho} (Published Unexamined Patent Applications)
and
{\it Tokkyo Koho} (Published Examined Patent Applications).
These gazettes are digitized,
but not organized because they do not trace changes in trade name
or firms' addresses.
To solve these problems,
Tamada et al. has organized a database \cite{Tamada06} and
this paper is based on theirs.
It includes 4,998,464 patents
published from January 1994 to December 2003 in patent gazettes.

The industrial cluster program of Japan began in April 2003
but preparatory steps had not been done until March 2005.
Hence, these patent data indicate
a situation where the industrial cluster program had not yet affected firms.
This means that we can study
the innate characteristics of firms without the program having an effect,
and discuss a preferable plan as to
how to take advantage of the characteristics.

\begin{figure}[bt]
\begin{center}
\includegraphics[scale=0.5]{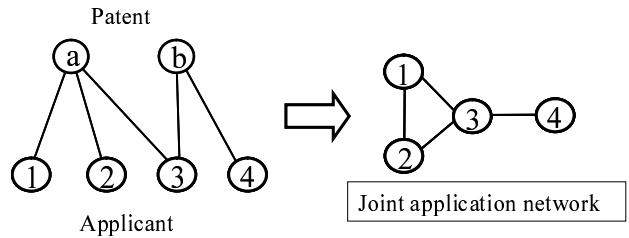}
\caption{Examples outlining how a joint application network is created.
In the figure at left, there is patent `a' that
has applicants `1', `2', and `3' and `b' that has `3' and `4'.
Nodes are applicants in a joint application network.
Applicants are connected to each other if they apply for a patent together.
There is a joint application network
in the figure at right.}
\label{fig:patentNetwork}
\end{center}
\end{figure}

We extracted applicants' data from the front pages of patents
and obtained a joint application network
(called a patent network after this).
A patent network has applicants as nodes, 
and joint applications as links (Fig. \ref{fig:patentNetwork}).
The links do not have weight or directions.
The applicants include firms and individuals.
However, the objective of this paper is 
to discuss what an industrial cluster should be.
We hence need a patent network that only consists of firms.
We consequently removed the nodes of individuals
and the links they had from the patent network.

How to develop a local network is a key issue
in discussing an industrial cluster.
However, analyses of the entire network can also contribute to this.
Therefore, 
we will discuss analyses of the entire network
in the rest of this section,
particularly, the basic properties of the patent network,
the degree distribution, the neighbors' average degree,
and the clustering coefficient.

The largest connected component is a part of the network
where all nodes can traverse each other,
and which has the largest number of nodes.
A network with all nodes has
67,659 nodes and 111,860 links,
and the largest connected component has
34,830 nodes and
84,843 links, which represent approximately 51\% and 76\% 
of the network with all nodes, respectively.

Here, let us give a definition of measurements.
The degree is the number of links a node has.
If a node, {\it i}, has {\it k} links,
the degree of node {\it i} is {\it k}.
The clustering coefficient \cite{Watts98}
is the measurement of triangles a node has.
Node {\it i}'s clustering coefficient
is quantified by $C_{i}=2e_{i}/k(k-1)$ where
$k$ is the degree and
$e_{i}$ is the number of links connecting the $k$ neighbors to each other.
The path length is the minimum number of links we need to travel between two nodes.

The average path length is 4.45 and
the longest path length is 18 
in the largest connected component.
The average path length and the longest path length
are based on all the combination of nodes in the network.
We will discuss
the possibility of reducing these path lengths
in the latter part of this section.
The clustering coefficient of the network with all nodes
is 0.29 and one of the largest connected component is 0.31.
The clustering coefficient here is the average of all nodes included
in the network.
The clustering coefficient of other firms' networks are smaller than this.
For example, the clustering coefficient of a firms' transaction network
is 0.21 \cite{Souma06}.
Generally, patent-network links are sparse compared to
those for other networks
because joint applications cannot occur without other linkages,
such as alliances, and transactions.
This means firms especially tend to form groups in a patent network.
We will take a closer look at the clustering coefficient
in the latter part of this section.

\begin{figure}[bt]
\begin{center}
\includegraphics[scale=0.45]{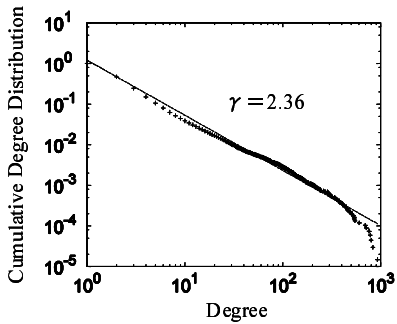}
\caption{Degree distribution.
Broken line is given by 
$p(k)\propto k^{-\gamma}$ with $\gamma=2.36$.}
\label{fig:degreeDistribution}
\end{center}
\end{figure}

Figure \ref{fig:degreeDistribution} plots
the degree distribution of the network of all nodes.
The horizontal axis represents the degree, $k$,
and 
the vertical axis indicates the cumulative distribution,
which is given by $P(k)=\int_{k}^{\infty}dk\ p(k)$,
where $p(k)$ means the degree distribution.
If $p(k) \propto k^{-\gamma}$,
$P(k\leq) \propto k^{-(\gamma-1)}$.
The broken line in the figure
is given by $p(k)\propto k^{-\gamma}$ with $\gamma=2.36$.
From the line, we can see that
the degree distribution follows a power law.
This means that the patent network can be categorized as a scale-free network.

The average path length
in a scale-free network
follows $ \mbox{log}\,\mbox{log}\,N$, 
where $N$ is the number of nodes.
This means that nodes can reach many other nodes by traveling a few steps.
If a node is connected to nodes that can reach in two links
in the patent network,
the average degree is 104.7 times larger than the one of the original network.
Creating new links is an important issue
to promote industrial clusters.
Since, intuitively,
connecting firms that do not have any relation is difficult,
this result means that
connecting firms connected to the same hub (a node with a large degree) 
can provide good opportunities.

\begin{figure}[bt]
  \begin{tabular}{ccc}
    \begin{minipage}{0.45\textwidth}
      \begin{center}
	\includegraphics[scale=0.45]{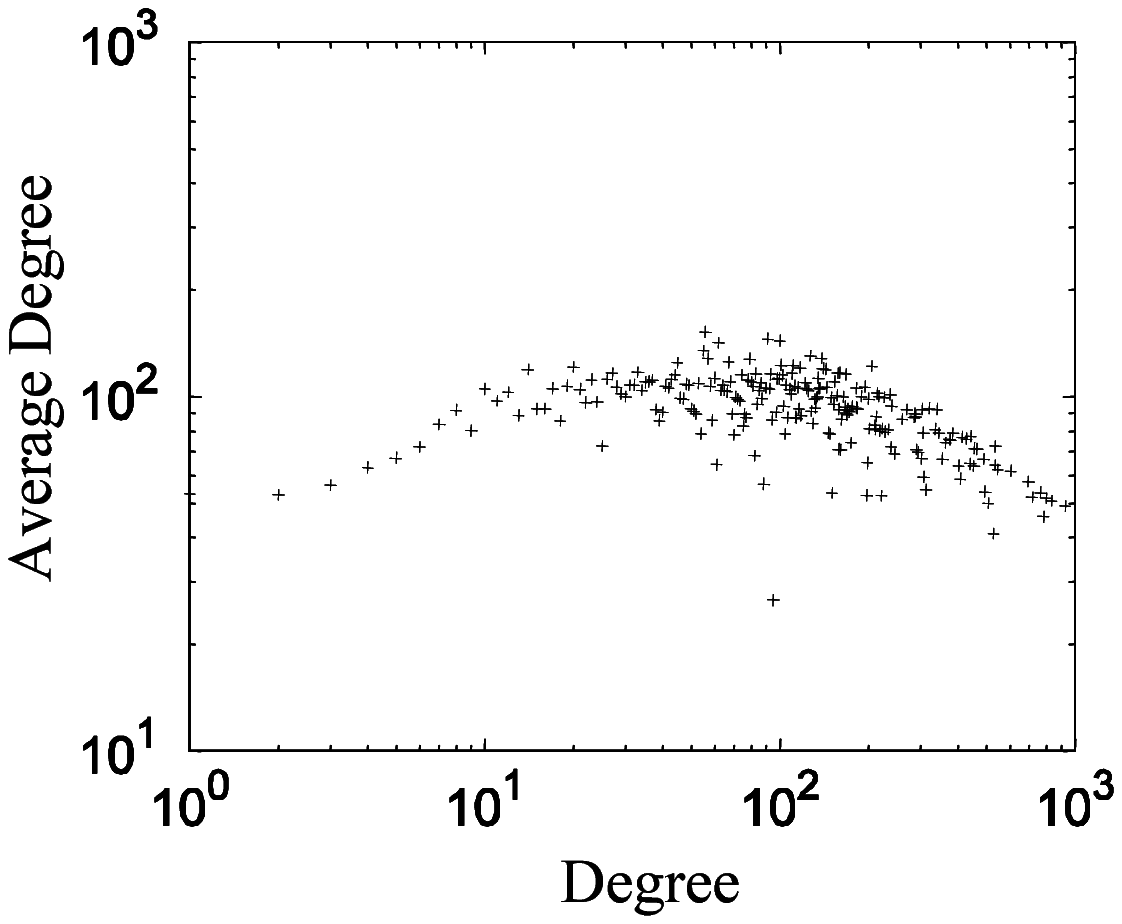}
	\caption{Average degree $<\!k_{nn}\!>$.
This figure shows nearest neighbors' average degree of nodes
with degree $k$.}
	\label{fig:degreeCorrelation}
      \end{center}
    \end{minipage} &
    \begin{minipage}{0.1\textwidth}
    \end{minipage} &
    \begin{minipage}{0.45\textwidth}
      \begin{center}
	\includegraphics[scale=0.45]{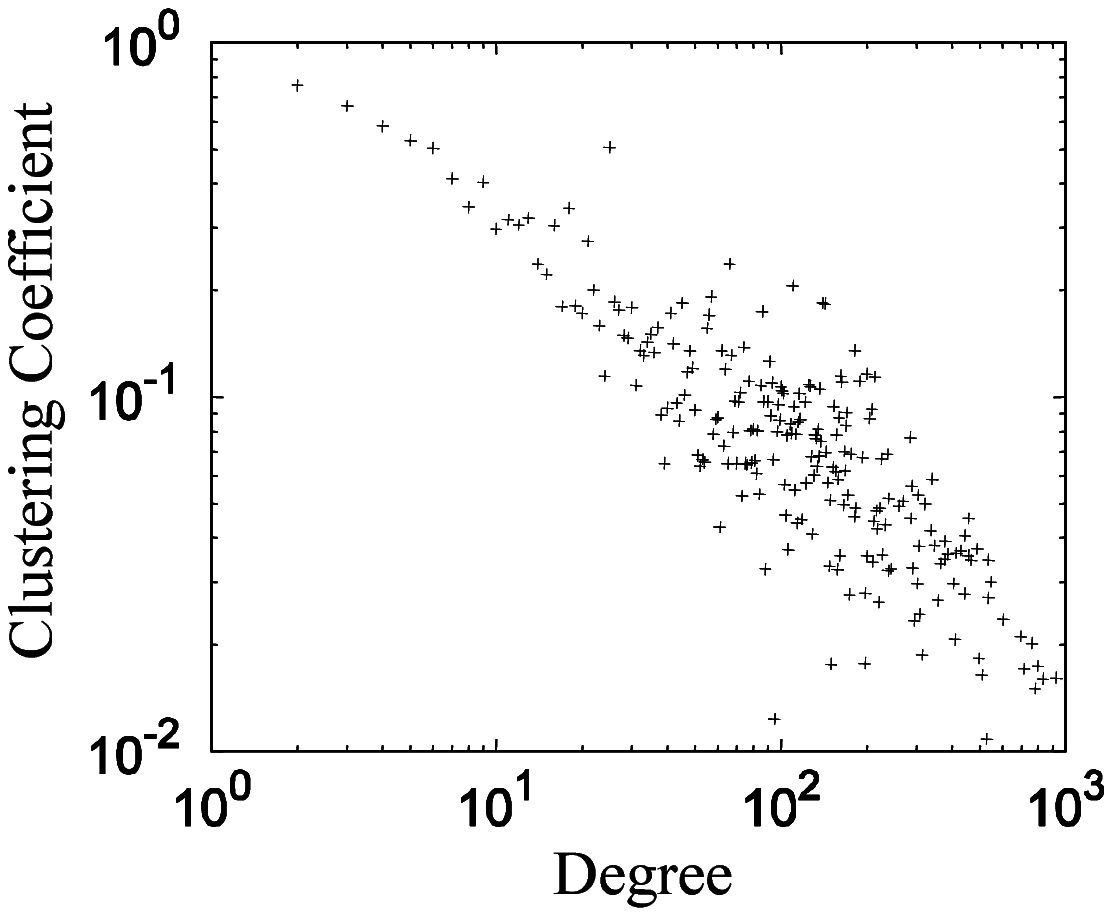}
	\caption{Clustering coefficient $C(k)$. This figure shows
average clustering coefficient of all nodes with degree $k$.}
	\label{fig:degreeClustering}
      \end{center}
    \end{minipage}
  \end{tabular}
\end{figure}

We will next discuss the neighbors' average degree.
Calculating the degree correlation 
is one of useful measurements
to discuss a network structure \cite{Pastor-Sattorass02}.
This is represented by
conditional probability $p_c(k'|k)$,
the probability
that a link belonging to a node with degree $k$
will be connected to a node with degree $k'$.
If this conditional probability is
independent of $k$, 
the network has a topology without any correlation between the nodes'
degree.
That is, $p_c(k'|k) = p_c(k') \propto k'p(k')$.
In contrast,
the explicit dependence on $k$ reveals
nontrivial correlations between nodes' degree,
and the possible presence of a hierarchical structure in the network.
The direct measurement of the $p_c(k'|k)$ is
a rather complex task due to the large number of links.
More useful measurement is
\mbox{$<\!k_{nn}\!>=\sum_{k'}{k'p_{c}(k'|k)}$},
i.e., the neighbors' average degree of nodes with degree $k$.
Figure \ref{fig:degreeCorrelation} shows
the average degree of the network of all nodes.
The horizontal axis shows the degree $k$,
and the vertical axis shows 
the degree correlation $<\!k_{nn}\!>$.
Figure \ref{fig:degreeCorrelation}
has a null correlation.
Therefore, we cannot see
the presence of a hierarchical structure for the degree in the network.
If a hierarchical structure exists,
there is the possibility
that we can reduce the length between nodes by creating a shortcut.
From the point of view of industrial clusters,
reducing the length between firms is
important for creating new links.
However, this result for the neighbors' average degree
denies such a possibility.

We will now discuss the clustering coefficient.
Its definition has already been given,
and the average clustering coefficient for all nodes has also been presented.
Figure \ref{fig:degreeClustering}
shows the average clustering coefficient
for the network of all nodes with degree $k$.
The horizontal axis indicates degree $k$,
and the vertical axis represents
the average clustering coefficient, $C(k)$, of all nodes with degree $k$.
There seems to be a negative correlation.
If a scale-free network has $C(k) \propto k^{-1}$,
the network is hierarchically modular \cite{Barabasi04}.
This hierarchical structure is different to the one of degree,
which was previously discussed.
The hierarchical structure here means that
sparsely connected nodes are part of highly clustered areas,
with communication between different highly clustered
neighbors being maintained by a few hubs.
This means that
a node may have to traverse redundant hubs
to access nodes in other modules.
Reducing the length between firms
in industrial clusters
is important for creating new links.
Therefore,
bridging firms in different modules directly
may positively affect the interactions between them.

\section{Spatial characteristics of patent networks}
\label{sec:spatial}

A lot of countries regard the regional accumulation
of firms' interactions as important
in policies of industrial clusters.
This section discusses the spatial characteristics
of the patent network.

\begin{figure}[bt]
\begin{center}
\includegraphics[scale=0.55]{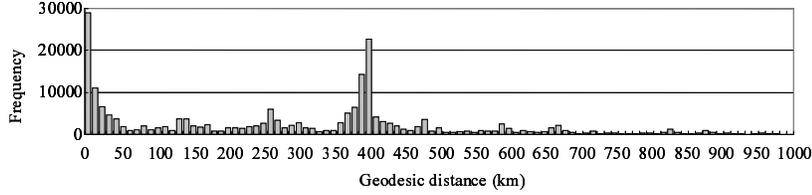}
\caption{Frequency of distance.
Distance between each link is calculated by nodes' addresses
connected by link.
There are several peaks in this chart.}
\label{fig:distance}
\end{center}
\end{figure}

Figure \ref{fig:distance} plots
the frequency of distance for all links.
The horizontal axis represents the geodesic distance, and
the vertical axis indicates the frequency.
The distance between each link is based on the nodes' addresses
connected by the link.
The addresses are converted to pairs of
latitudes and longitudes,
and the geodesic distance is calculated from these.

There are several peaks.
The largest one is around the first 10 km.
The peak means that the nearer the firms,
the more likely they are to have links.
This supports the assumption of policies on industrial clusters
because regional accumulation can be seen as a natural behavior of firms.
However, there are other peaks.
These are around 130 km, 250 km, and 400 km.
The cause of these peaks seems to be that
there are cities connected by {\it Shinkansen} (bullet train).
Japan has a well organized infrastructure for public transportation,
hence cities at long distances can have many links.
This indicates that
geographic adjacency is not the exact reason for
the facilitation of links,
but the time it takes to travel.
If it takes a short time to travel between firms,
they are likely to have links.
Consequently,
the infrastructure for public transportation should be considered
to discuss industrial clusters at least in Japan.

\begin{figure}[bt]
\begin{center}
\includegraphics[scale=0.45]{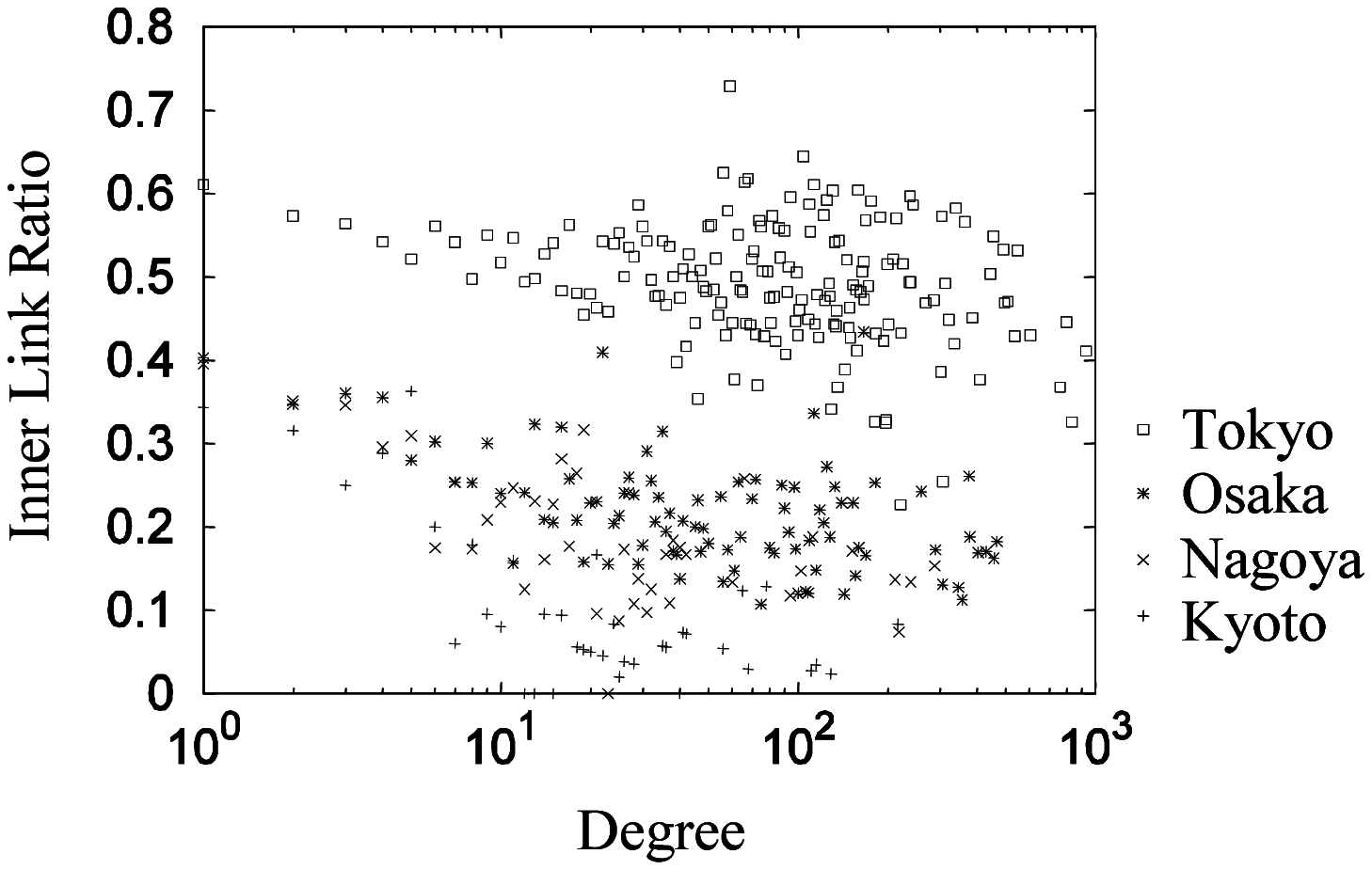}
\caption{Average link ratio, $\overline{r_k}$.
There are negative correlations in each city.
}
\label{fig:innercity}
\end{center}
\end{figure}

We will now define the inner link ratio of a node in a specific city as
$r_k=k_{inside}/k$, where 
$k$ is degree, and $k_{inside}$ is
the number of links connected to nodes in the same city.
The average link ratio, $\overline{r_k}$, is
an average of $r_k$ over all nodes in the same city with degree $k$.
We picked out four cities that
include numerous firms.
They were Tokyo 23 special-wards
(a primary area of the capital), Osaka, Kyoto,
and Nagoya.
Figure \ref{fig:innercity} shows the average link ratio for all four cities.
The horizontal axis indicates the degree $k$,
and the vertical axis represents
the average link ratio, $\overline{r_k}$.

There are negative correlations in all four cities.
This means that
a node with a small degree prefers to have links with nodes in the same region.
However,
a node with a large degree
tends to have links with nodes in other regions.
It is thus not appropriate to adhere
to create links among firms in the same region
in a discussion on industrial clusters,
and we should consider
links that firms already have.

Summing up,
the patent network reveals valuable indications of industrial cluster policies.
The indications can be summarized as follows.
(1) A firm can be connected to many nodes
via hubs.
(2) Bridging different modules may shorten the paths between nodes in them.
(3) The distance between links reveals the importance of the time it takes
to travel.
(4) We have to consider the existing links firms have
to facilitate the creation of new links.








\end{document}